\documentstyle[12pt,fleqn]{article}
\newfont{\bigssq}{cmr10 scaled\magstep5}

\catcode`\@=11

\begin {document}
\title{\bf \ Stability of Naked Singularity arising in gravitational
collapse of Type I matter fields}

\author {Sanjay B.~Sarwe \footnote{email:sbsarwe\_ngp@sancharnet.in}\\
  Department of Mathematics, S. F. S. College,
  \\ Seminary Hill,
   Nagpur-440 006, India  \\
 R. V. Saraykar  \footnote{email:sarayaka\_ngp@sancharnet.in} \\
 Department of Mathematics, Nagpur University \\
 Campus, Nagpur-440 033, India}
\date{}
 \maketitle

  \begin{abstract}
 Considering gravitational collapse of Type I matter fields, we
prove that, given an arbitrary $C^{2}$- mass function
$\textit{M}(r,v)$ and a $C^{1}$- function $h(r,v)$ (through the
 corresponding $C^{1}$- metric function $\nu(t,r)$),
there exist infinitely many choices of energy distribution
function $b(r)$ such that the `true' initial data
($\textit{M},h(r,v)$) leads the collapse to the formation of naked
singularity. We further prove that the occurrence of such a naked
singularity is stable with respect to small changes in the initial
data. We remark that though the initial data leading to both black
hole and naked singularity form a ``big" subset of the true
initial data set, their occurrence is not generic. The terms
`stability' and `genericity' are appropriately defined following
the theory of dynamical systems. The particular case of radial
pressure $p_{r}(r)$ has been illustrated in details to get clear
picture of how naked singularity is formed and how, it is stable
with respect to initial data.
\end{abstract}
\hspace{.4in} Keywords: {Gravitational collapse,
central-singularity, stability}

\hspace{.3in} PACS: {04.20.-q, 04.20.Dw, 04.40.Dg }

\section {Introduction}
In recent papers \cite {jd, jo, jr} Joshi, Dwivedi and Goswami
have discussed the role of initial data in spherically symmetric
gravitational collapse for Type I matter fields. It is shown that
given the density ($\rho$) and pressure profiles ($ p_{r},
p_{\theta} $) at the initial spacelike hypersurface from which the
collapse evolves, there is a wide choice available to choose the
velocity function and rest of the initial data so that the end
state of collapse is either a black hole (BH) or a naked
singularity (NS). This result is significant for two reasons:- (1)
it produces a substantially ``big'' initial data set which under
gravitational collapse results into a naked singularity. (2) Type
I matter fields include most of the known physical forms of matter
like dust, perfect fluid. At least in spherically symmetric case,
this result shows that while considering black hole physics, one
must discard certain part of initial data set that leads the
collapse to a naked singularity. However, all of the initial data
is not independent, so it poses a question, which initial data set
shall evolve into a NS as a result of gravitational collapse?
Further, having a NS for certain initial data set is not enough
because if it looses its characteristic for small perturbation in
the neighbourhood of that certain initial data set then the NS is
not serious enough to challenge cosmic censorship conjecture \cite
{rp}. Therefore, the all important question ie. is the NS
developed from certain initial data set a stable one?

This motivates us to study both the above mentioned questions. We
prove that given a $C^{2}$-mass function $M(t,r)$ and
 $C^{1}$- velocity function $\nu(t,r)$ on an initial spacelike
hypersurface, there always exists a $C^{1}$- function $ b(r)$ such
that the total initial data comprising of mass function (includes
density and pressures), a function $\nu(r)$ and energy
distribution function $b(r)$ evolves into a naked singularity as a
result of gravitational collapse.

  F.I. Cooperstock and et al. \cite {cjjs} have shown that
if the radial pressure $p_{r}$ is positive , $r>0$ and having
strictly positive mass function then non-central singularity is
covered by the horizon irrespective of the sign of tangential
pressure $p_{\theta}$. The stability of central shell-focusing
singularity at $r=0$ that we are discussing in this paper is with
respect to small changes in the initial data comprising of mass
function and velocity function. "Small" changes in the initial
data are with respect to appropriate norm on the the functions
space of all physically reasonable initial data. This concept of
stability is analogous to that of structural stability in the
theory of dynamical systems and is explained in section 4.
Essentially, we prove that the set of initial data leading the
collapse to central shell focusing naked singularity is an open
subset of the set of all initial data in a suitable functions
space. That we get a covered singularity which is non-central when
$p_{r} > 0$ does not disturb the stability of central shell
focusing singularity (naked or covered) with respect to small
changes in initial data.

In section 2, we briefly summarize  analysis given in \cite {jr}
and state the conditions on the initial data under which the
collapse will lead to a naked singularity. In section 3, we
analyze the condition of existence of naked singularity to prove
our assertion mentioned above by using existence theory of first
order ordinary differential equations. Thereafter, in section 4,
we define and explain the concept of stability and genericity that
we follow in this paper and we prove openness of initial data set.
We, then, discuss genericity of this set. In section 5, we study
the particular case $p_{r} (r)$ to demonstrate the involvement of
$p_{r}$ and $p_{\theta}$ as an initial data in the occurrence of
NS, its stability and non-generic nature. Finally, we make some
concluding remarks.

\section {Field equations, initial data and naked singularity}
In this section, we summarize  analysis from \cite {jr}. The
general spherically symmetric metric describing space-time
geometry within the collapsing cloud can be described in comoving
coordinates $(t, r, \theta, \phi)$ by
\begin{equation}
 ds^2 = - e^{2 \nu(t,r)}dt^2 +  e^{2 \psi(t,r)}dr^2 + R^2(t,r) d
\Omega^2 \label {m01}
\end{equation}
where $ d\Omega^2=d\theta^2+ \sin^2 \theta \; d \phi^2 $ is the
metric on an $2$-sphere. The stress-energy tensor for Type I field
in a diagonal form is  given by \cite {he}
\begin{equation}
 T{^{t}_{t} } = -\rho, T{^{r}_{r} } = p_{r},
 T{^{\theta}_{\theta}}=T{^{\phi}_{\phi} } = p_{\theta} \label {em03}
\end{equation}
The quantities $\rho$, $p_{r}$ and $p_{\theta}$ are the energy
density, radial and tangential pressures respectively. We take the
matter field to satisfy weak energy condition i.e. the energy
density measured by any local observer be non-negative, so for any
vector $V^{i}$, we must have, $T_{ik}V^{i} V^{k} \geq 0$ which
means ${{\rho}{\geq}}0$;  ${\rho+p_{r}{\geq}} 0$;  ${\rho +
p_{\theta}{\geq}} 0$.

Einstein field equations for the metric (1) are
\begin{equation}
\rho = \frac{F'}{R^{2}R'} \label {fe04}
\end{equation}
\begin{equation}
 p_{r} = - \frac{\dot{F}}{R^{2} \dot{R}} \label {fe05}
 \end{equation}
\begin{equation}
\nu' = \frac{2(p_{\theta}-p_{r})}{\rho + p_{r}}\frac{R'}{R} -
\frac{p'_{r}}{\rho + p_{r}} \label {fe06}
\end{equation}
\begin{equation}
 - 2 \dot{R'} + R'\frac{\dot{G}}{G} + \dot{R}
\frac{H'}{H} = 0 \label {fe07}
\end{equation}
\begin{equation}
\hspace{.5in} G - H = 1 - \frac{F}{R} \label {fe08}
\end{equation}
where we have defined $ G(t,r) = e^{-2 \psi } (R')^2 $ and $
H(t,r) = e^{ -2 \nu }( \dot{R})^2$.

The total mass in a shell of collapsing cloud of comoving radius
$r$ is represented by the arbitrary function $ F= F(t,r) $. The
weak energy conditions imply $ F \geq 0 $. The regularity at the
initial epoch $t=t_{i}$ is preserved by $F(t_{i},0) = 0$ i.e. the
mass function should vanish at the centre of the cloud. The
density singularity in the space-time is at $R = 0$ and $R' = 0$,
the later one is due to shell-crossings and can possibly be
removed from the space-time \cite {r1}, so we consider here only a
physical singularity where all matter shells collapse to a zero
physical radius known as shell-focusing singularity. We can use
scaling freedom available for the radial co-ordinate $r$ to
introduce the function $v(t,r)$ by the relation
\begin{equation}
R(t,r) = r v(t,r), \label {sf09}
\end{equation}
we have $v(t_{i},r) = 1$; $ v(t_{s}(r), r) = 0 $ and for collapse
$ \dot{v} < 0 $. Above relation is obtained by defining $ v(t,r) =
R / r $ \cite {rg2}. The time $ t = t_{s}(r) $ corresponds to the
shell-focusing singularity $ R = 0$. The six arbitrary functions
of the shell radius $r$ as given by $ \nu(t_{i},r) = \nu_{o}(r)$,
\ $\psi(t_{i},r) = \psi_{o}(r)$, \ $R(t_{i},r) = r$, \
$\rho(t_{i},r) = \rho_{o}(r)$, \ $p_{r}(t_{i},r) = p_{r_{o}}(r)$,
 \ $p_{\theta}(t_{i},r) = p_{\theta_{o}}(r)$ evolve the dynamics of
the initial data prescribed at the initial epoch $ t = t_{i}$.
From equation (\ref {fe06}), we obtain
\begin{equation}
\nu_{o}(r)=\int_{0}^{r} \left( \frac{2 (p_{\theta_{o}} -
p_{r_{o}})} {r ( \rho_{o} + p_{r_{o}})} -
\frac{p'_{r_{o}}}{\rho_{o} + p_{r_{o}} }\right)dr  \label {ie10}
\end{equation}
which indicates that not all the above initial data are mutually
independent and that $ \nu_{o}(r)$ has the form,
\begin{equation}
\nu_{o}(r) = r^2 g(r) \label {v11}
\end{equation}
where $g(r)$ is at least a $C^{1}$ function of $r$ for $r = 0$,
and at least a $C^{2}$ function for $r>0$. Let us now assume the
physically reasonable behaviour of the initial data at the center
$ r=0$ i.e. $p'_{r_{o}}(0) = p'_{\theta_{o}}(0) = 0$ and
$p_{r_{o}}(0) - p_{\theta_{o}}(0) = 0 $. For details, we refer
\cite {jd}. We have a total of five equations with seven unknowns,
namely $ {\rho}, {p_{r}}, {p_{\theta}}, {\nu}, {\psi}, R $   and
$F$ giving us the freedom of choice of two functions. Selection of
these two free functions, subject to the given initial data and
the weak energy condition above, determines the matter
distribution and the metric of the space-time and thus, leads to a
particular time evolution of the initial data. The existence and
uniqueness of solution of the system of field equations with above
mentioned initial data has been discussed by Joshi and Dwivedi
\cite {dj, jd}. The solution continues to exist in the
neighbourhood of the singularity given by $R=0$. A general mass
function for the cloud can be considered as
\begin{equation}
F(t,r) = r^3 \textit{M}(r,v) \label {m12}
\end{equation}
where $\textit{M}(r,v)$ is regular and continuously twice
differentiable. Using equation (\ref {m12}) in equations (\ref
{fe04}) and (\ref {fe05}), we obtain
\begin{equation}
\rho = \frac{3\textit{M} + r [ \textit{M}_{, r} + \textit{M}_{, v}
 \ v']}{ v^2 ( v + r v')}; \hspace{.1in}  \ p_{r} = -
\frac{\textit{M}_{, v}}{v^2} \label {m13}
\end{equation}
Then as $ v \rightarrow 0 \; , \rho \rightarrow \infty $ and
$p_{r}\rightarrow \infty $ i.e. both the density and radial
pressure blow up at the singularity. The regular density
distribution at the initial epoch is given by $\rho_{o}(r) =
3\textit{M}(r,1) + r {\textit{M}(r,1)} _{,\ r}$ .

We take the initial surface to be the cloud given by $ 0 \leq r
\leq r_{c}$ for some finite $r_{c}$ on which the initial data,
namely the mass function $F(t,r)$, the metric function $ \nu
(t,r)$ and the function $b(r)$ ( to follow) evolve as the collapse
begins according to the Type I field equations. So, we take
${\textit{}{M}(r,v)} > 0 $ and at least $C^{2}$ in ${\mathcal{D}}
\equiv [0,r_{c}]\times [0,1]$, for varying $ r \in [0, r_{c}], v
\in [0, 1] $.
 Consider the general
 metric function,
\begin{equation}
\nu(t,r) = A(t, R) \label {cv14}
\end{equation}
where $A(t, R)$ is arbitrary, continuously differentiable function
of $t$ and $R$, with the initial constraint $ A(t_{i},R) =
\nu_{o}(r) $. We use equation (\ref {cv14}) in equation (\ref
{fe06}) to get

\begin{equation}
2 p_{\theta} = R A_{,R}( \rho + p_{r}) + 2 p_{r} + \frac{R p'_{r}}
{R'}   \label {tp15}
\end{equation}
  From  above equation, we conclude that the tangential pressure
also blows up at the singularity. Also, the use of equation (\ref
{cv14}) in equation (\ref {fe07}), yields
\begin{equation}
G(t,r) = d(r) e^{2(A - \int A_{,t} dt) } \label {g16}
\end{equation}
where $ d(r) $ is another arbitrary continuously differentiable
function of $r$. From equation (\ref {v11}), we generalize the
form of  $ A(t, R) $ as $ A(t, R) = r^2 g_{1}(r, v)$ , where $
g_{1}(r,v)$ is a continuously differentiable function and $
g_{1}(r,1) = g(r) $. Similarly, we have $ A - \int A_{,t} dt = r^2
g_{2}(r,v) $ and at the initial epoch $ g_{2}(r,1) = g(r)$ . We
write
\begin{equation}
d(r) = 1 + r^2 b(r).  \label {b17}
\end{equation}
where b(r) is the energy distribution function for the collapsing
shells. Then using equations (\ref {m12}),(\ref {cv14}) and (\ref
{g16}) in equation (\ref {fe08}), we get
\begin{eqnarray}
 \sqrt{R} \dot{R} = - e ^{r^2 g_{1}(r,v)} \
 {\sqrt{[ 1 + r^2 b(r)] R e^{r^2 g_{2}(r,v)} - R + r^3 \textit{M}
}}  \hspace{.2in} \label {rd18}
\end{eqnarray}
where negative sign is chosen since, for the collapse, $ \dot{R} <
0 $. Defining a function $h(r,v)$ as
\begin{equation}
 h(r,v) = \frac{e^{r^2 g_{2}(r,v)} - 1}{r^2} = 2
g_{2}(r,v)+ O(r^2v^2)   \label {h19}
\end{equation}
and substituting equation (\ref {h19}) in equation (\ref {rd18}),
we get
\begin{equation}
 \sqrt{v} \dot{v} = - \sqrt{e^{2 r^2 (g_{1} + g_{2})}v b(r)
+ e^{2r^2 g_{1} } ( v h + \textit{M})}. \label {v20}
\end{equation}
Integrating the above equation , we have
\begin{equation}
 t(v,r) = \int_{v}^{1} \frac{ \sqrt{v} dv}{\sqrt{e^{2 r^2
(g_{1} + g_{2})}v  b(r) + e^{2r^2 g_{1} } ( v h + \textit{M})}}
\hspace{.4in} \label {i21}
\end{equation}
Here, the variable $r$ is treated as a constant in the above
equation. Expanding $t(v,r)$ around the centre, we get
\begin{equation}
t(v,r) = t(v,0) + r \chi(v) + O(r^2) \label {t22}
\end{equation}
where the function
\begin{equation}
\chi(v) = - \frac{1}{2} \int_{v}^{1} {  \frac{\sqrt{v} [ v b'(0) +
v h_{,r}(0,v) + \textit{M}_{,r}(0,v)]}{[ v b(0) + v h(0,v) +
\textit{M}(0,v)]^{3/2}}} dv.  \hspace{.2in} \label {ch23}
\end{equation}

Now, it is clearly seen that the value of $\chi(0)$ depends on the
functions $b(0),\textit{M}(0,v)$ and $ h(0,v)$, which in turn,
depend on the initial data at the initial surface $ t=t_{i}$.
Thus, a tangent to the singularity curve $ t = t_{s_{o}} $ is
completely determined by the given set of density, pressure,
velocity function $\nu$ and function $b(r)$. Further, from
equation (\ref {v20}), we can write
\begin{equation}
\sqrt{v} v' = \chi(v) \sqrt{ v b(0)  + v h(0,v) + \textit{M}(0,v)}
+ O(r^2)  \label {st24}
\end{equation}

If the neighbourhood of the centre $ R = 0, r = 0$ gets trapped
earlier than the formation of singularity, then it is covered (
i.e. occurrence of a black hole), and if otherwise happens the
singularity is naked (ie. singularity can be observed locally or
by a faraway observer) \cite{r1}. For examination of the nature of
central singularity at $ R = 0, r = 0$, we consider the equation
for outgoing radial null geodesic, $ dt/dr = e^{(\psi - \nu )} $
and test, if there would be any families of null geodesics coming
out of the singularity. Further, we write the null geodesic
equation in terms of the variables ( $ u= r^{\alpha}, R $),
choosing ($ \alpha = 5/3 $ ), and using equation (\ref {fe08}), we
obtain
\begin{equation}
  \frac{dR}{du} = \frac{3}{5} \left( \frac{R}{u} +
\frac{\sqrt{v}v'}{\sqrt{\frac{R}{u}}} \right) \left( \frac{1 -
\frac{F}{R}}{ \sqrt{G}[ \sqrt{G} + \sqrt{H}]} \right)  \label
{rn25}
\end{equation}
If the null geodesics terminate at the singularity in the past
with a definite tangent, then at the singularity, we have $
 dR/du > 0 $, in the $ (u, R) $ plane with a finite value.
Hence, all points $r> 0$ on the singularity curve are covered
since $ F/R \rightarrow \infty $ with $  dR/du \rightarrow
{-\infty}$ and only the singularity at the centre $ r=0$ could be
naked. For the case, when $R' > 0$ near the central singularity,
we have
\begin{equation}
x_{o} = \lim_{t\rightarrow t_{s}} \lim_{ r\rightarrow {o}}
\frac{R}{u} =  \frac{dR}{du} \Big{|} _{t\rightarrow
t_{s},r\rightarrow {o}} \label {pr26}
\end{equation}
and   use of equations (\ref {st24}) and (\ref {rn25})
 yield  $ {x_{o}}^{3/2} = (3/2) \sqrt{\textit{M}(0,0)}
\chi(0) $ and the radial null geodesic emerging from the
singularity in $(R,u)$ co-ordinates is $ R = x_{o} u$, or in
$(t,r)$ plane, it is given by $ t - t_{s}(0) = x_{o} r^{5/3}$.
Therefore, if $ \chi(0) > 0 $ then $ x_{o} > 0$, and we get
radially outgoing null geodesics coming out from the singularity ,
giving rise to a naked central singularity. However, if $ \chi(0)
< 0 $ then   we have a black hole solution, as there will be no
such trajectories coming out. For $ \chi(0) = 0 $ , we will have
to take into account the next higher order non-zero term in the
singularity curve equation, and a similar analysis can be carried
out by choosing a different value of $\alpha$.

  Thus, it is clearly seen as to how the initial data determines
the NS/BH phases as end states of collapse, since $ \chi(0)$ is
determined by these initial profiles for the collapsing matter
given by equation (\ref {ch23}).

\section {Existence of energy distribution function $ b(r)$ leading to NS}
In this section, we prove the assertion mentioned in the
introduction. We choose $b(r)$ to satisfy the differential
equation on a constant v-surface
\begin{equation}
 \frac{1}{2}  \frac{\sqrt{v} [v b'(r)  + v h_{,r}(r,v) +
\textit{M}_{,r}(r,v)] }{[ v b(r)  + v h(r,v) +
\textit{M}(r,v)]^{3/2}} = B(r,v)  \label {as27}
\end{equation}
for $ \ 0 \leq r \leq r_{c} \ $ where  $ B(r,v) $ is a continuous
function defined on $ {\mathcal{D}}$ such that
\begin{equation}
B(0,v) = \frac{1}{2} \frac{\sqrt{v} [ b'(0) v + v h_{,r}(0,v) +
\textit{M}_{,r}(0,v)]} {[ v b(0)  + v h(0,v) +
\textit{M}(0,v)]^{3/2}} < 0   \label {ba28}
\end{equation}
for all $ v $ in $[0,1]$. It will then follow that
\begin{equation}
\chi(0) = \lim_{v \rightarrow {0}} \chi(v) = - \int_{0}^{1} B(0,v)
dv > 0 . \label {nc29}
\end{equation}
This condition ensures that central shell-focusing singularity
will be naked.

We, now, discuss the existence of $b(r)$ as a solution of the
differential equation (\ref {as27}). We put
\begin{equation}
x(r,v) = v b(r) + v h(r,v) + \textit{M}(r,v)   \label {sb30}
\end{equation}
a continuous function of $r$, in equation (\ref {as27}) and
rearranging it, we get
\begin{equation}
\frac{dx}{dr} = \frac{1}{\sqrt{v}} \left[ 2 B(r,v) x^{3/2}
  \right] \equiv f(x,r) \label {de31}
\end{equation}
with the initial condition
\begin{equation}
 x(0,v) = v b(0) + v h(0,v) + \textit{M}(0,v)  \label {ic32}.
\end{equation}
Let us ensure the existence of $ C^{1} $ - function $x(r,v)$ as a
solution of above initial value problem defined throughout the
cloud. The function $f(x,r)$ is continuous in $r$, with $x$
restricted to a bounded domain. With such domain of $r$ and $x$,
$f(x,r)$ is also  $ C^{1} $ - function in $x$ which means $f(x,r)$
is Lipschitz continuous in $x$. Therefore, the differential
equation (\ref {de31}) has a unique solution satisfying initial
condition (\ref {ic32}).

  Further, we can ensure that the solution will be defined over the
entire cloud i.e. for all $r$ in $ [0,r_{c}]$ by using the freedom
in the choice of arbitrary function $B(r,v)$. For this , we
consider  the domain $[0,r_{c}] \times [0,d]$ for some finite $d$.
Let $S$ be the supremum of the modulus of $f(x,r)$. Then the
differential equation (\ref {de31}) has a unique solution defined
over the entire cloud provided \cite{ve}
\begin{equation}
r_{c} \leq \inf (r_{c}, \frac{d}{S}) = \frac{d}{S}  \label {ec33}.
\end{equation}
This yields
\begin{equation}
\max_{{0\leq r \leq r_{c}},   {0 \leq x \leq d} } \hspace{.1in}
\left| \frac{1}{\sqrt{v}} \left[ 2 B(r,v) x^{3/2}  \right] \right|
\leq \frac{d}{r_{c}} \label {ma34}.
\end{equation}
Condition (\ref {ec33}), in turn, will be satisfied if the weaker
condition
\begin{equation}
0 \leq |B(r,v)| \ x^{3/2} \leq \frac{d \ \sqrt{v}}{2 \ r_{c}}
\label {sm35}
\end{equation}
holds for all $r$ in $[0,r_{c}]$.

The collapsing cloud may start with  $r_{c}$ small enough so that
the expression \ $
   d \sqrt{v}/{2r_{c}}$
    \ which is always positive, satisfies the condition (\ref
{sm35}) with $x$ restricted to a bounded domain. We then have
infinitely
 many choices for the function
$B(r,v)$, which is continuous and satisfies conditions (\ref
{ba28}) and (\ref {sm35}) for each choice of $v$. For each such
$B(r,v)$, there will be a unique solution $x(r,v)$ of the
differential equation (\ref {de31}), satisfying initial condition
(\ref {ic32}), defined over the entire cloud and in turn, there
exist a unique function $b(r)$ for each such choice of $B(r,v)$,
that is given by the expression
\begin{equation}
b(r) = \frac{ x(r,v) -\textit{M}(r,v) - v h(r,v) }{v} \label {h36}
\end{equation}
over $[0,r_{c}]$. Thus, for a given constant $v$-surface and given
initial data of mass function $ F(t,r) = r^{3}\textit{M}(r,v) $
and $ h(r,v) = (e^{r^2 g_{2}(r,v)} - 1) /{r^2} $ satisfying
physically reasonable conditions (expressed on $\textit{M}$),
there exists infinitely many choices for the function $b(r)$   \
such that condition (\ref{ba28}) is satisfied. The condition
continues to hold as $v \rightarrow 0$, because of continuity.
Hence, the central singularity developed in the collapse is a
naked singularity.

\section {Definition of stability and genericity of NS/BH}
In this section, we give the appropriate definitions of stability
and genericity. Our definitions of stability and genericity are
based on the structural stability of a dynamical system, and
genericity of a property of a dynamical system respectively. For
more details, refer chapter 7, \S3,4 of Abraham and Marsden
\cite{am}.

Let $\textbf{M}$ be a manifold on which a vector field or a
dynamical system is defined. Let $\mathcal{X} (\textbf{M})$ be the
space of all vector fields on $\textbf{M}$. Whitney
$C^{r}$-topology on $\mathcal{X} (\textbf{M})$, generated by the
norm is given by

\begin{eqnarray*}
 \parallel f \parallel_{r} =
 \sup \left\{ \sum_{k=0}^{r}{\parallel D^{k}f(u) \parallel}/u \in U
 \right\},
\end{eqnarray*}
where $V$ and $W$ are vector spaces, $U$ an open subset of $V$,
and $f: U \rightarrow W$. $D^{k} f$ denotes $k$th Frechet
derivative of $f$. $\mathcal{X} (\textbf{M}) $ endowed with
Whiteny $C^{r}$- topology is denoted by $ {\mathcal{X}}^{r}
(\textbf{M})$.

Let $X_{1}$ be a vector field or a dynamical system on
$\textbf{M}$. Then $X_{1}$ is structurally stable if there is a
neighbourhood $\Phi$ of $ X_{1} \in {\mathcal{X}}^{r}(\textbf{M})$
in the Whitney $C^{r}$- topology such that $ Y_{1} \in \Phi$
implies $X_{1}$ and $Y_{1}$
 are topologically conjugate i.e. they have equivalent phase portraits.
 This means that there is a homeomorphism $ h_{o}: \textbf{M} \rightarrow \textbf{M}$
 carrying oriented orbits of $X_{1}$ to oriented orbits of $Y_{1}$.

A property of vector fields in $ {\mathcal{X}}^{r} (\textbf{M})$
is a proposition $P(x_{1})$ with a variable $ x_{1} \in
{\mathcal{X}}^{r}(\textbf{M})$. A property $P(x_{1})$ with a
variable $ x_{1} \in {\mathcal{X}}^{r}(\textbf{M})$ is generic if
the subset $ \{ x_{1} \in {\mathcal{X}}^{r}(\textbf{M}) / P(x_{1})
\} \subset {\mathcal{X}}^{r}(\textbf{M}) $ contains a residual
set.

A subset $A$ of a topological space $X_{1}$ is called residual if
and only if $A$ is the intersection of a countable family of open
dense subsets of $X_{1}$. A topological space $X_{1}$ is a Baire
space if and only if every residual set is dense. We also know
that every complete metric space and in particular every Banach
space is a Baire space. Also, whether $ \textbf{M}$ is compact or
not, ${\mathcal{X}}^{r}(\textbf{M})$ is a Baire space.

In our case, we apply these definitions analogously to the
evolution of initial data into a gravitational collapse leading to
a naked singularity or a black hole. We treat evolving initial
data as a vector field or a dynamical system and consider the
space of all initial data with sufficient differentiability
defined on a collapsing compact spherical shell of Type I  field,
in place of $\mathcal{X}(\textbf{M})$, endowed with a suitable
$C^{r}$-topology. Property $P$ of a dynamical system becomes the
property of initial data, namely, whether this initial data leads
the collapse to a naked singularity or a black hole. Thus, the
definitions of stability of a naked singularity and genericity of
its occurrence can be stated as follows.

Let $I_{o}$ be the initial data set, which when evolves, leads the
collapse to a NS/BH. We say that a NS/BH is stable, if there is a
neighbourhood I of $I_{o}$ in $C^{r}$-topology such that, if
$I_{1}$ is another initial data in I, then $I_{1}$ also leads the
collapse to a NS/BH. In other words, if the set of initial data
leading the collapse to a NS/BH forms an open subset of
$C^{r}$-space of all initial data, then the NS/BH will be stable.
Similarly, occurrence of a NS/BH will be said to be generic, if
the set of all initial data leading to NS/BH, is a dense subset of
the parent $C^{r}$-space.
 All sets of initial data are required to satisfy the constraint
equations, in addition to the energy conditions.

\subsection {Stability of Naked Singularity}
The analysis in section III shows that the only conditions on the
initial data which evolves the collapse into a naked singularity
are the energy conditions. Hence, for stability, we have to
examine energy conditions only. What we now show is that, the set
of such initial data satisfy energy conditions forms an open
subset of the space of all initial data. For this , we use the
technique of Saraykar and Ghate \cite{sg}. The energy conditions
${{\rho}{\geq}}0$; ${\rho+p_{r}{\geq}} 0$; ${\rho +
p_{\theta}{\geq}} 0$ can be written in the form
\begin{eqnarray*}
\left[ 3 \textit{M} + r \textit{M}_{,r} + r v'
\textit{M}_{,v}\right] \equiv E_{1} \geq 0, \hspace{.4in}
 \left[ 3
\textit{M} + r \textit{M}_{,r} - v \textit{M}_{,v}\right] \equiv
E_{2} > 0 , \nonumber  \\
 \left[3 w \textit{M} + r w \textit{M}_{,r} + w_{1} \textit{M}_{,v}
- r v (v+rv') \textit{M}_{,rv} \right] \equiv E_{3} \geq 0
\end{eqnarray*}
 where \ $w = 2(v + r v') + r v \nu_{,r}$
and \ $w_{1} = 2 r^{2} v'^{2} -(2 + r) v^{2} \nu_{,r}$ \ on the
domain  $ {\mathcal{D}}$.
 Here, we have used an assumption
that $v$ is an increasing function of $r$ so that $v'\geq 0$ on
any surface $ t = t_{j}$ say, so that no shell-crossing
singularity condition $(ie. R' > 0)$ holds. So, this is a valid
assumption. We know that as the nuclear fuel exhausts in a star,
it starts collapsing under its tremendous  gravitational pull, and
in this case, in the formation of central singularity radial
pressure diverges, so we assume radial pressure $p_{r}$ to be
non-negative throughout the gravitational collapse. Therefore,
from equation (\ref{m13}), we have $\textit{M}_{,v} \leq 0$. Then,
$ E_{1} \leq E_{2} $. Hence, only $ E_{1} $ and $E_{3}$ shall take
part in further stability analysis.

We assume that $X$ be an infinite dimensional Banach space of all
$C^{2}$ real-valued functions defined on $ {\mathcal{D}}$, endowed
with the norms
\begin{eqnarray*}
\parallel \textit{M}(r,v) \parallel_{1} = \sup_{{\mathcal{D}} } |\textit{M}  |
+ \sup_{{\mathcal{D}} } |\textit{M}_{,r}| + \sup_{{\mathcal{D}} }
|\textit{M}_{,v}|    \hspace{.8in} \nonumber \\  and
\parallel \textit{M}(r,v) \parallel_{2} = \sup_{{\mathcal{D}} } |\textit{M}  |
+ \sup_{{\mathcal{D}} } |\textit{M}_{,rr}  |  +
\sup_{{\mathcal{D}} } |\textit{M}_{,rv}|  + \sup_{{\mathcal{D}} }
|\textit{M}_{,vv} |
\end{eqnarray*}
These norms are equivalent to the standard $C^{1}$ and $C^{2}$
norms
\begin{eqnarray}
\parallel \textit{M}(r,v) \parallel_{1} = \sup_{{\mathcal{D}} } (|\textit{M} |
+ |\textit{M}_{,r}| + |\textit{M} _{,v}|) \hspace{.6in} \
\nonumber
\\  and
 \parallel \textit{M}(r,v) \parallel_{2} = \sup_{{\mathcal{D}}
} (|\textit{M} | + |\textit{M}_{,rr}| + |\textit{M} _{,rv}|+
|\textit{M}_{,vv}|)  \label {def37}
\end{eqnarray}
Let ${\mathcal{G}} = \{ \textit{M}(r,v):\textit{M}> 0, \textit{M}
$ \ is $ \ C^{2},  \ E_{1} > 0 $ \ and $ E_{3} > 0$ on $
{\mathcal{D}} \}$ \ be a subset of $X$.

We show that ${\mathcal{G}}$ is an open subset of $X$. For
$\textit{M}$ in ${\mathcal{G}}$, let us put $\delta =
\min(\textit{M})$, $\gamma = \min( E_{1})$ , $\beta = \min( E_{3}
)$, $\lambda = \max(w)$, $\lambda_{1} = \max(w_{1})$, $\lambda_{2}
= \max(v + r v')$ and $\lambda_{3} = \max(v')$ for varying $r$ in
$[0,r_{c}]$ and $ v \epsilon [0,1]$, the functions involved herein
are all continuous functions defined on a compact domain
${\mathcal{D}}$ and hence, their maxima and minima exist.
 We define a positive real number
 \begin{eqnarray*}
  \mu = \frac{1}{2} \min\{\delta , \frac{\gamma}{9},
\frac{\gamma}{3 r_{c}}, \frac{\gamma}{3 r_{c} \lambda_{3}}, \frac{
\beta} {12 \lambda}, \frac{ \beta} {4 r_{c} \lambda}, \frac{
\beta} {4 \lambda_{1}}, \frac{ \beta} {4 r_{c} \lambda_{2}}\} .
\end{eqnarray*}
Let $\textit{M}_{1}(r,v)$ \ be  \ $C^{2}$ \ in \ $ {\mathcal{D}}$
\ with \ $\parallel \textit{M} - \textit{M}_{1}
\parallel_{1}  < \mu$ and $\parallel \textit{M} - \textit{M}_{1}
\parallel_{2}  < \mu$.
Using definition (\ref {def37}), we get $| \textit{M}_{1} -
\textit{M} | < \mu$  \ , $| \textit{M}_{1,r} - \textit{M}_{,r} | <
\mu$  \ , $| \textit{M}_{1,v} - \textit{M}_{,v} |< \mu$ and $|
\textit{M}_{1,rv} - \textit{M}_{,rv} | < \mu$ over ${\mathcal{D}}
$. Therefore, for choice of $\mu$, the respective inequalities are
\begin{eqnarray}
\textit{M}_{1} > \textit{M} - \frac{\delta}{2} > 0 , \hspace{.1in}
3 |\textit{M}_{1} - \textit{M} | < \frac{\gamma}{6} ,
\nonumber \\
 r |\textit{M}_{1,r} - \textit{M}_{,r} | \leq r_{c}
|\textit{M}_{1,r} - \textit{M}_{,r} | < \frac{\gamma}{6} ,
\hspace{.1in}
 r v' | \textit{M}_{1,v} - \textit{M}_{,v} |  \leq r_{c} \lambda_{3}
 | \textit{M}_{1,v} - \textit{M}_{,v} | < \frac{\gamma}{6}, \nonumber \\
3 w |\textit{M}_{1} - \textit{M} | < \frac{\beta}{8} ,
\hspace{.2in}
 r w |\textit{M}_{1,r} - \textit{M}_{,r} | \leq r_{c}
\lambda |\textit{M}_{1,r} - \textit{M}_{,r} | < \frac{\beta}{8} ,
\nonumber  \\
 w_{1} | \textit{M}_{1,v} - \textit{M}_{,v} | \leq
r_{c} \lambda_{1} | \textit{M}_{1,v} - \textit{M}_{,v} | <
\frac{\beta}{8}, \hspace{.7in}
 r v ( v + r v') | \textit{M}_{1,rv}
- \textit{M}_{,rv}|<\frac{\beta}{8}  \label {de38}
 \end{eqnarray}
 that are satisfied on ${\mathcal{D}}$.
 The  $II^{nd}$ , $III^{rd}$ and $IV^{th} $ inequations from above
 yield
\begin{eqnarray*}
 3 |\textit{M}_{1} - \textit{M} | + r
|\textit{M}_{1,r} - \textit{M}_{,r} | + r v'|\textit{M}_{1,v} -
\textit{M}_{,v} |< \frac{\gamma}{2} < {\gamma}  \leq E_{1}
\hspace{.2in}
\end{eqnarray*}
Further, we can write \hspace{.1in} \ $|[3 \textit{M}_{1} + r
\textit{M}_{1,r}+ r v' \textit{M}_{1,v}] - E_{1}| < E_{1} $ \
where $ E_{1} > 0 $ on $ {\mathcal{D}}$.
 Hence, $ [3 \textit{M}_{1} + r \textit{M}_{1,r}+ r v' \textit{M}_{1,v}] > 0 $ on $
{\mathcal{D}}$. Using similar analysis for last four inequations
of equation (\ref{de38}), we obtain \ $ [3 w \textit{M}_{1} + r w
\textit{M}_{1,r} + w_{1} \textit{M}_{1,v} - r v (v+rv')
\textit{M}_{1,rv}] > 0$ on $ {\mathcal{D}} $ \ provided $w > 0$
and $w_{1} > 0$ on ${\mathcal{D}}$.

 Thus,\ $ \textit{M}_{1}  > 0$,
$\textit{M}_{1}$ is $C^{2}$ , $ [3 \textit{M}_{1} + r
\textit{M}_{1,r}+ r v' \textit{M}_{1,v}] > 0 $ and $ [3 w
\textit{M}_{1} + r w \textit{M}_{1,r} + w_{1} \textit{M}_{1,v} - r
v (v+rv') \textit{M}_{1,rv}] > 0$ on $ {\mathcal{D}} $ \ provided
$w > 0$ and $w_{1} > 0$  throughout ${\mathcal{D}}$.
 Therefore, $\textit{M}_{1}(r,v)$ also
lies in ${\mathcal{G}}$ \ and hence,\ ${\mathcal{G}}$ is an open
subset of $X$.

The conditions $w>0$ and $w_{1}>0$ required for openness of
${\mathcal{G}}$ on ${\mathcal{D}}$, impose conditions on the
function $\nu(t,r)$, $\nu(t,r) = A(t, R)$ and $ A - \int A_{,t} dt
= r^2 g_{2}(r,v) $, therefore, $h(r,v) = \ e^{r^2 g_{2}(r,v) - 1}
/ {r^2}$ is constrained by the conditions $w>0$ and $w_{1}>0$ in
 domain ${\mathcal{D}}$. Now, let $Y$ be an infinite dimensional
Banach space of all $C^{1}$ real-valued functions defined on
${\mathcal{D}}$. The set ${\mathcal{H}}$ of functions $h(r,v)$
with restrictions $w>0$ and $w_{1} > 0$ would form (by similar
technique) an open subset of $Y$. Then ${\mathcal{G}} \times
{\mathcal{H}}$ is an open subset in the product space $X \times
Y$.

 Taking $(\textit{M}_{1}, h_{1}(r,v))$ in the neighbourhood of
$(\textit{M},h(r,v))$ in $G \times {\mathcal{H}}$, and using
equation (\ref {sm35}) analogously for $ \textit{M}_{1}$ and
$h_{1}$, we have a choice of infinitely many $B_{1}(r,v)$, such
that for each such $B_{1}(r,v)$, there will exist a unique
$b_{1}(r)$ so that the initial data of mass function $r^{3}
\textit{M}_{1}$ and $h_{1}(r,v)$ together will lead the collapse
to formation of a naked singularity. Thus, naked singularity
arising from $(\textit{M},h(r,v))$ is $C^{2}$ - stable in the
sense defined above.

The analysis given above, guarantee the existence of a metric
function $ \nu(t,r)$ for a given initial data set. Such choice of
$\nu(t,r)$ and expressions for $G$ and $H$ together will yield the
metric (\ref {m01}) as an exact solution leading to the occurrence
of naked singularity. Thus, given initial data and the weak energy
conditions above determine the matter distribution and the metric
of the space-time.

\subsection {Genericity of NS/BH}
Let us discuss the genericity of the occurrence of these
singularities. We have seen that in Type I matter field collapse,
the end state of collapse is governed by the choice of initial
data. As explained above, occurrence of a NS is generic if the set
${\mathcal{G}} \times {\mathcal{H}} $ of `true' initial data (
\textit{M}, h(r,v) ), contains a residual set in the space. The
space $X \times Y$ contains negative functions which cannot be
realized as a limit of a sequence of positive functions belonging
to ${\mathcal{G}} \times {\mathcal{H}}$. Hence,
${\mathcal{G}}\times {\mathcal{H}} $ is not dense in $X \times Y$.
The space under consideration is a Banach space. Every complete
metric space and in particular every Banach space is a Baire
space, and since in a Baire space, every residual set is dense, it
follows that ${\mathcal{G}} \times {\mathcal{H}} $ is not
residual. Therefore, we conclude that occurrence of naked
singularities is not $C^{2}$-generic. Further, since the set
${\mathcal{G}} \times {\mathcal{H}}$ is open, it cannot be nowhere
dense and hence ${\mathcal{G}}\times {\mathcal{H}} $ is not a
meagre set either. Thus, it is a substantially ``big" set.

\section {Special case $p_{r}=p_{r}(r)$ }
The case $p_{r} = 1$ has been studied by Goswami and Joshi  and it
is shown that the presence of a non-vanishing pressure gradient
gives rise to either formation of NS or a BH {\cite {jr1}}. We
illustrate above analysis by studying a special case where $p_{r}$
be a function of $r$ alone. It gives a more clear picture of the
occurrence of naked singularity and its stability. We choose two
allowed free functions $p_{r}$ and $\nu(t,r)$ as follows:
\begin{equation}
p_{r} = p_{r}(r), \hspace{.1in} \nu(t,r) = c(t) + {\eta} (R)
\label {p40}
\end{equation}
Now, let us see, how the Einstein's field equations react to this
choice which will decide the evolution of collapse. As $p_{r}$ is
a function $r$ alone, integrating equation (\ref{fe05}), we obtain
\begin{equation}
F(t,r) = - \frac{p_{r}}{3} R^{3} + z(r) \label {f41}
\end{equation}
where $z(r)$ is another arbitrary function of $r$. We need to
choose $z(r)$, $ z(r) = \frac{2}{3} r^{3} p_{r}(r) $ so that the
mass function takes the form
\begin{equation}
F = \frac{p_{r}}{3}( 2 r^{3} - R^{3} ) \label {f42}
\end{equation}
that satisfies the regularity conditions at the initial epoch $
F(t_{i}, r) =  p_{r} r^{3}/3 $ and $ F(t_{i}, 0) = 0 $. Therefore,
$ F \geq 0 $ provided $ p_{r} \geq 0 $. Hence, non-negativity of
radial pressure shall maintain non-negativity of mass function
throughout the gravitational collapse. Next, equation (\ref{fe04})
becomes
\begin{equation}
\rho = \frac{1}{3} \ \frac{3 p_{r} ( 2 r^{2} - R^{2} R')
  + p'_{r}  \ (2 r^{3} - R^{3})}{ R^{2} R'} \label {d43}
\end{equation}
The density blows up at the singularity in the limit of approach
to the singularity ie. as $ r \rightarrow 0 $ and $ t \rightarrow
t_{s}$. At the initial epoch $ t = t_{i}$,
\begin{equation}
 \rho = p_{r} + \frac{1}{3} r p_{r}'  \geq 0   \label {id44}
\end{equation}
 , an all important relation between energy
density and radial pressure at the initial epoch.
\\
Equation (\ref{fe07} ) takes the form
\begin{equation}
 G = e^{2 \ \eta(R)} d_{p}(r)  \label {g45}
\end{equation}
with $d_{p}(r) = 1 + r^{2} b_{p}(r)$ where $b_{p}$ is the energy
distribution function for the collapsing shells. Using equation
(\ref{p40}) in equation (\ref{fe06}), we obtain
\begin{equation}
( \rho + p_{r} ) \ \eta_{,R} \ R' R  = 2 ( p_{\theta} - p_{r} ) R'
- R \ p'_{r}  \label {r46}
\end{equation}
Since the density blows up at the singularity, from above
equation, it is clear that the tangential pressure also blows up
in the limit of approach to the singularity. Further, we find that
all of initial data is not independent at the initial epoch $t =
t_{i}$
\begin{equation}
( \rho + p_{r} ) \ \eta_{,R} \ r  = 2 ( p_{\theta} - p_{r} ) - r \
p'_{r} \  \label {p47}.
\end{equation}
Above equation and equation (\ref{id44}) suggest that  $b_{p}$,
radial and tangential pressures will be sufficient to form initial
data set for the collapse. We are going to show the existence of
$b_{p}$ through $ p_{r} $ and $ p_{\theta}$. Therefore, we
consider $p_{r},p_{\theta}$ as an
 initial data set. Using this initial data prescribed at the initial epoch $ t = t_{i}$
 to evolve the collapse, we
 can integrate the equation (\ref{r46}) and obtain,
 \begin{equation}
\eta(R) =  \int_{0}^{R} \frac{2 (p_{\theta_{o}} - p_{r_{o}}) - R
p_{r}' } {R ( \rho_{o} + p_{r_{o}})}dR   \label {v48}
\end{equation}

Thus, velocity distribution function $ \eta(R)$ is determined by
smooth functions $ p_{r}$ and $ p_{\theta}$. Using equations
(\ref{f42}) and (\ref{g45}), equation (\ref{fe08}) can be written
as
\begin{eqnarray}
 \sqrt{R} {\dot{R}} = - \ a(t) \ e^{\eta} \
   { \sqrt{[ 1 + r^2 b_{p}] Re^{2 \eta}- R  +
\frac{p_{r}}{3} ( 2 r^3 - R^{3}) }}   \label {d49}
\end{eqnarray}
where $a(t)$ is a function of time and we choose $a(t) = 1$ by
suitable scaling of time co-ordinate.\\
We define
\begin{equation}
h_{p}(R) = \frac{ e^{2 \eta(R)} - 1} {R^2} = 2 \ g_{p}(R) + O(R^2)
\label {d50}
\end{equation}
Using this definition, equation (\ref{d49}) takes the form
\begin{equation}
    \sqrt{v} \ {\dot{v}} = -
   \sqrt{v b_{p} e^{4 \eta} + e^{2 \eta}[ v^{3} h_{p}(rv) +
\frac{p_{r}}{3} ( 2 - v^{3})] }  \label {v51}
\end{equation}
Integrating above equation, we obtain
\begin{equation}
 t(v,r) = \int_{v}^{1} \frac{\sqrt{v} dv} {\sqrt{v e^{4
\eta} b_{p} +  e^{2\eta} [ v^{3} h_{p}(rv) + \frac{p_{r}}{3} ( 2 -
v^{3})]} }   \label {i52}
\end{equation}
where the variable $r$ is treated as a constant. Further, the time
taken for the central shell to reach the singularity is given by
\begin{equation}
 t_{s_{o}} = \int_{0}^{1} \frac{\sqrt{v} dv} {\sqrt{v b_{p}(0)
+ v^{3} h_{p}(0) + \frac{p_{r}}{3}(0) ( 2 - v^{3})} } \label
{ts53}
\end{equation}
and $t_{s_{o}}$ is well defined provided $ [ \ v b_{p}(0) + v^{3}
h_{p}(0) + \ {p_{r}(0)} ( 2 - v^{3}) /3 \ ] > 0 $. The time taken
for other shells to reach the singularity can be determined from
the Taylor expansion of $ t(v,r) $ around $r=0$ as $ t\rightarrow
t_{s}$,
\begin{equation}
t_{s}(r) = t_{s_{o}} + r \chi_{p}(v) + O(r^2) \label {tsr54}
\end{equation}
where $\chi_{p}(v) = \ {dt(v,r)/{dr}} \big{|}_{r=0}$ and
\begin{eqnarray}
 \chi_{p}(v) = - \ {\frac{1}{2}} \
       \int_{v}^{1} {  \frac{\sqrt{v}
[ v b_{p}'(0) + v^{4} h_{p}{,_{r}}(0) + \frac{p_{r}'(0)} {3}(2-
v^{3})]}{[ v b_{p}(0) + v^{3}   h_{p}(0) + \frac{p_{r}(0)} {3}
(2-v^{3}) ]^{3/2}}} dv. \label {cho55}
\end{eqnarray}
For any constant v surface, we have $d(v) = 0$ which implies \
$\sqrt{v} \ \dot{v} = - \ (1 / \chi_{p}(v)) \sqrt{v} \ v'$.
 Now, equation (\ref {v51}) takes the form
\begin{equation}
\sqrt{v} v' = \chi_{p}(v)
   \sqrt{v  b_{p}(0)+  v^{3} h(0) +
\frac{p_{r}(0)}{3} ( 2 - v^{3}) }  + O(r^2)  \label {cv56}
\end{equation}
The radial null geodesic equation (\ref {rn25}) in the variables
$(u = r^{\alpha}, R)$ as $ t\rightarrow t_{s}$, $ r \rightarrow
0$, gives
\begin{equation}
x_{o}^{3 / 2} = \frac{3}{2} \chi_{p}(0)  \sqrt{ \frac{2} {3}
p_{r}(0)  } \label {rd57}
\end{equation}
From equation (\ref{cho55}), it is clear that $ \chi_{p}(v)$
depends on the functions $ b_{p}(0), h_{p}(0), p_{r}(0)$ and these
functions, in turn, depend on the given set of density, pressure
and velocity function $ \nu$. Therefore, equation (\ref{rd57})
depicts that tangent to the singularity curve $ t = t_{s_{o}}$ is
completely determined by the given set of initial data and radial
pressure in particular. The radial null geodesic emerging from the
singularity in $(R,u)$ co-ordinates is $ R = x_{o} u$, while in
$(t,r)$ plane, the null geodesic near the singularity is given by
$ t - t_{s}(0) = x_{o} r^{5/3}$.

It is clear that if $ \chi_{p}(0) > 0 $ then $x_{o} > 0 $ and with
this, we get radially outgoing null geodesics coming out from the
singularity, giving rise to a naked central singularity. However,
no such trajectory comes out if $ \chi_{p}(0) < 0 $ as we have a
black hole situation. Thus, we have some region of space-time for
which $ \chi_{p}(0) > 0 $ depending on initial data, leading to
occurrence of central naked singularity in that space-time region.
Therefore, we choose $b_{p}$ to satisfy the following differential
equation on a constant $v$-surface
\begin{eqnarray}
  \frac{1}{2} \frac{\sqrt{v} [ v b_{p}'(r) + v^{4} h_{p}{,_{r}}(r v)
   + \frac{p_{r}'(r)} {3}(2- v^{3})]}
  {[ v b_{p}(r) + v^{3} h_{p}(r v) + \frac{p_{r}(r)} {3} (2-v^{3}) ]^{3/2}}
  dv
   = { \mathcal{B}}(r,v)
 \label {ex58}
\end{eqnarray}
where $ { \mathcal{B}}(r,v)$ is a continuous function defined on
${\mathcal{D}}$ such that
\begin{eqnarray}
  \frac{1}{2} \frac{\sqrt{v} [ v b_{p}'(0) + v^{4} h_{p}{,_{r}}(0)
  + \frac{p_{r}'(0)} {3}(2- v^{3})]}
  {[ v b_{p}(0) + v^{3} h_{p}(0) + \frac{p_{r}(0)} {3} (2-v^{3}) ]^{3/2}}
  dv
    = { \mathcal{B}}(0,v) < 0 \hspace{.1in}  for\ all \hspace{.1in}  v \ \epsilon \ [0,1].
 \label {exp59}
\end{eqnarray}
It then follows that
\begin{equation}
\chi_{p}(0) = \lim_{v \rightarrow {0}} \chi_{p}(v) = -
\int_{0}^{1} { \mathcal{B}}(0,v) \ dv > 0 . \label {ep60}
\end{equation}
The above requirement ensures the nakedness of central shell
focusing singularity. Now, we discuss existence of $b_{p}$ as a
solution of the differential equation (\ref{ex58}). For this, we
use
\begin{equation}
y(r,v) =  v \ b_{p}(r) + v^{3} h_{p}(r v) + \frac{p_{r}(r)} {3}
(2-v^{3})   \label {sub61}
\end{equation}
a continuous function of $r$, in equation (\ref{ex58}), and we
obtain
\begin{equation}
\frac{dy}{dr} = \frac{1}{\sqrt{v}} \left[ 2 { \mathcal{B}}(r,v)
y^{3/2}
  \right] \equiv \varphi(y,r) \label {de62}
\end{equation}
with initial condition
\begin{equation}
 y(0,v) = v b_{p}(0) + v^{3} h_{p}(0) + \frac{p_{r}(0)} {3}
(2-v^{3})  \label {ic63}.
\end{equation}
This initial value problem defined throughout the cloud  has a
 $C^{1}$ - function $y(r,v)$ as its unique solution defined over
 the entire cloud. Using the arguments of general case
 (section III) for the function $ \varphi(y,r)$ and $ { \mathcal{B}}(r,v)$ )
 a weaker condition
\begin{equation}
0 \leq |{ \mathcal{B}}(r,v)| \ y^{3/2} \leq \frac{ n \sqrt{v}}{2 \
r_{c}} \label {co64}
\end{equation}
holds for all $r$ in $[0,r_{c}]$ and $v \in [0,1]$ where $n$ is a
finite number such that $0 \leq y \leq n$. This condition holds
true for $y$ restricted to a bounded domain and the collapse may
start with $r_{c}$ small enough. Therefore, there are infinitely
many choices for the function ${ \mathcal{B}}(r,v)$, which is
continuous and satisfies the conditions (\ref{exp59}) and
(\ref{co64}) for each choice of $v$. For each such ${
\mathcal{B}}(r,v)$, there will be a unique solution $y(r,v)$ of
the differential equation (\ref{de62}), satisfying initial
condition (\ref{ic63}), defined over the entire cloud. This unique
solution $y(r,v)$ for each such choice of ${ \mathcal{B}}(r,v)$
yields
\begin{equation}
b_{p} = \frac { y(r,v) - v^{3} h_{p}(rv) - \frac{p_{r}(r)} {3}
(2-v^{3})} {v} \label {bp65}
\end{equation}
over $[0, r_{c}]$. As each function appearing in the above
expression is continuous, therefore, the condition holds true as $
v \rightarrow 0$, because of continuity. We know that velocity
distribution function $ \eta(R)$ is determined through smooth
functions $ p_{r}$, $ p_{\theta}$ on a given constant $v$-surface,
which in turn, through equation (\ref{d50}) determines the
function $h_{p}(rv)$. Thus, for a given constant $v$-surface and
given initial data of radial and tangential pressures $ p_{r}$ and
$ p_{\theta}$, there exists infinitely many choices of the
function $b_{p}$ such that condition (\ref{exp59}) is satisfied.
We have thus proved the existence of energy distribution function
$b_{p}$ for a given set of initial data leading to the occurrence
of a naked central shell focusing singularity.

Now, we illustrate the stability of occurrence of NS. For
simplicity, we consider the evolution of initial data on the
initial surface $t=t_{i}$ where $v=1$ and $v'=0$. Using equations
(\ref{id44}) and (\ref{p47}), the energy conditions
${{\rho}{\geq}}0$; ${\rho+p_{r}{\geq}} 0$;  ${\rho +
p_{\theta}{\geq}} 0$ can be described as
\begin{eqnarray*}
\rho = p_{r} + \frac{1}{3} r p_{r}' \equiv K_{1} \geq 0,
\hspace{.3in}
 2p_{r}+ \frac{1}{3} r p_{r}'  \equiv K_{2} \geq 0,  \nonumber  \\
2 (2 + r \eta_{,R} ) p_{r}  + [ 1 + \frac{1} {3} (2 + r \eta_{,R}
) ] r p_{r}' \equiv K_{3} \geq 0
\end{eqnarray*}
We denote by $X_{p}$ be an infinite dimensional Banach space of
all $C^{1}$ real-valued functions defined on $ [0,r_{c}]$, endowed
with the $C^{1}$ norm. Let us take \
 $G_{p} = \{ p_{r}(r):p_{r}> 0,
p_{r} $  is  $  C^{1}, K_{1} > 0 $ \ and \ $ K_{3} > 0$ on \ $
[0,r_{c}]$ \} be a subset of \ $X_{p}$.

Now, to show that $G_{p}$ is an open set in $X_{p}$. For $p_{r}$
in $G_{p}$, we put $\alpha_{1} = \min(p_{r})$, \ $\alpha_{2} =
\min( K_{1})$, \ $\alpha_{3} = \min( K_{3} )$, and $\lambda_{p} =
\max(2 + r \ \eta_{,R})$.
 We define a positive real number
 \begin{eqnarray*}
  \mu = \frac{1}{2} \min\{\alpha_{1} , \frac{\alpha_{2}}{3},
\frac{\alpha_{2}}{ r_{c}}, \frac{\alpha_{3}}{6 \lambda_{p}},
\frac{ \alpha_{3}} {3 r_{c}}, \frac{ \alpha_{3}} {3 r_{c}
\lambda_{p}} \}.
\end{eqnarray*}
Let $p_{r_{1}}(r)$ be $C^{1}$ in $ [0,r_{c}]$ with $\parallel
p_{r} - p_{r_{1}} \parallel  < \mu$. Now, using definition of
standard $C^{1}$ norm, we get $| p_{r} - p_{r_{1}} | < \mu$ \ and
$| p_{r}' - p_{r_{1}}' | < \mu$  over $ [0,r_{c}]$. Then, for each
choice of $\mu$, the respective inequalities
\begin{eqnarray}
p_{r_{1}} > p_{r} - \frac{\alpha_{1}}{2} > 0 , \hspace{.1in}
|p_{r_{1}} - p_{r}| < \frac{\alpha_{2}}{6} , \hspace{.1in}
\frac{1}{3} r |p_{r_{1}'} - p_{r}' | \leq \frac{1}{3} r_{c}
|p_{r_{1}}' - p_{r}' |
< \frac{\alpha_{2}}{6} ,\nonumber \\
2 (2+ r \eta_{,R}) |p_{r_{1}} - p_{r}| \leq 2 \lambda_{p}
|p_{r_{1}} - p_{r}| < \frac{\alpha_{3}}{6},  \hspace{.1in}
 r |p_{r_{1}'} - p_{r}' | \leq  r_{c}|p_{r_{1}}' - p_{r}' | <
\frac{\alpha_{3}}{6}, \nonumber   \\
\frac{r} {3} (2+ r \eta_{,R}) | p_{r_{1}}' - p_{r}' | \leq r_{c}
\lambda_{p} |p_{r_{1}}' - p_{r}' |< \frac{\alpha_{3}}{6} \nonumber
 \label {inq68}
\end{eqnarray}
 are satisfied in $[0,r_{c}]$. Using above inequalities analogous to
 the general case, we obtain
$p_{r_{1}} > 0, \ [ p_{r_{1}} + \ {r p_{r_{1}}'}/3 \ ]
> 0 $, \ $ [2 \ (2 + r \eta_{,R} ) p_{r_{1}}  + [ 1 + \ (2
+ r \eta_{,R} )/3 \ ] r p_{r_{1}}'] > 0$, provided $  (2 + r
\eta_{,R} ) > 0$ on $ [0,r_{c}]$. Therefore, $p_{r_{1}} \in G_{p}$
and hence, $G_{p}$ is an open set in $X_{p}$.

The function $\eta(R)$ is determined by $p_{r}$, $p_{\theta}$ on
the given constant $v$-surface and the condition $  (2 + r
\eta_{,R} ) > 0$ is necessary for openness of the set $G_{p}$.
Therefore, the $C^{1}$ function \ $h_{p}(R) = \ ( e^{2 \eta(R)} -
1) /{R^2}$ \ is also determined by initial data $p_{r}$,
$p_{\theta}$ on a given  constant $v$-surface constrained with $
(2 + r \eta_{,R} ) > 0$ on $ [0,r_{c}]$. Let ${\mathcal{H}}_{p}$
be a subspace of $X_{p}$ containing such initial data $h_{p}(R)$.
Then $G_{p} \times {\mathcal{H}}_{p}$ is an open set in the
product space \ $X_{p} \times X_{p} $.

Let $(p_{r_{1}}, h_{p_{1}}(r))$ be in the neighbourhood of
$(p_{r},h_{p}(r))$ in  $G_{p} \times {\mathcal{H}}_{p} $. Using
equation (\ref {co64}) analogously for $p_{r_{1}}$ and
$h_{p_{1}}(r)$, we have a choice of infinitely many ${
\mathcal{B}}_{1}(r,1)$, such that for each such ${
\mathcal{B}}_{1}$, there will exist a unique $b_{1}(r)$ so that
the initial data  $p_{r_{1}}$ and $h_{p_{1}}(r)$ together will
lead the collapse to formation of a naked singularity. Thus, naked
singularity arising from initial data $(p_{r},h_{p}(r))$ is
$C^{1}$ - stable. Thus, this particular case illustrates clearly
how the choice of initial data leads the collapse to a naked
singularity and how it is stable under small perturbations of
initial data in an appropriate mathematical sense.

\section {Discussion and conclusions}

1. Our main conclusions in the paper are the following: (a) Given
a $C^{2}$- mass function $ \textit{M}(r,v)$ and a $C^{1}$-
function $h(r,v)$, on any v = constant surface, we find a $C^{1}$-
energy distribution function $b(r)$ such that the collapse ends in
a naked singularity. (b) With physically reasonable conditions put
on mass function $\textit{M}$, and with $C^{1}$- function
$h(r,v)$, the initial data consisting of ($\textit{M},h(r,v)$)
leading the collapse to a naked singularity forms an open subset
of the space $X \times Y$. This establishes stability of naked
singularity with respect to initial data.

2. Similar analysis for the case $\chi(0)< 0$ ( that is $B(0,v)>
0$ for all $v$ in $[0,1]$) shows that the set $G \times
{\mathcal{H}}$ of `true' initial data ($\textit{M},h(r,v)$),
leading the collapse to a black hole forms an open set and it is
neither dense nor nowhere dense in the space of all initial data.
Thus, occurrence of black hole is stable but not generic.

3. Our analysis is carried out with given initial data on a
constant $v$- surface. Hence, the role of $\textit{M}$ and $b(r)$
can be interchanged (refer equation \ref {h36}) such that the mass
function would  evolve in the continuous gravitational collapse.

4. The conditions $w>0$ and $w_{1}>0$ which occurred in our
stability analysis of NS (general case) can yield some information
about the function $\nu$. For simplicity, let us consider these
conditions on the initial surface $t=t_{i}$. For $w>0$, we obtain
$\nu_{,r} > - 2/r$ for all $r \in [0, r_{c}]$. This means
$\nu_{,r} \in ( -2/r_{c} \ , \infty )$. Next, for $w_{1} > 0$, we
get $[ -(2 + r) \nu_{,r}] > 0$, this implies $ \nu_{,r} \in
(-\infty, 0)$. Thus, common domain of $\nu_{,r}$ is $( -2/r_{c} \
, 0)$. Therefore, $\nu$ is a decreasing function of $r$ in
$[0,r_{c}]$ for sufficiently small $r_{c}$.

While, in particular case $[2 + r \eta_{,R}] > 0$ yields $
\eta_{,r} \in (-2/r_{c} \ , \infty )$, therefore, $\eta(r)$ is a
decreasing function of $r$ in $[0,r_{c}]$. Also, the above said
condition can be written as  $[6 (p_{r} + p_{\theta})- r p_{r}'] >
0 $, it means $C^{1}$- stability of occurrence of NS is associated
with initial data $p_{r}$ and $p_{\theta}$ such that $[6 (p_{r} +
p_{\theta})- r p_{r}'] > 0 $. It is this choice of data that gives
rise to stable NS.

5. The space $X$ used in the analysis is an infinite-dimensional
Banach space, and as such, is not locally compact. It is not
therefore, straightforward to define a measure on this space.
Hence, we can not say whether the set ${\mathcal{G}}$ is of zero
measure or not. However, as mentioned earlier, ${\mathcal{G}}$ is
not meagre, and thus, it is a substantially `big' set. Similar
argument is applicable to the set ${\mathcal{G}} \times
{\mathcal{H}}$ in the product space $ X \times Y$. We believe that
rigorous analysis that we have worked out shall help further in
understanding the gravitational collapse of Type I matter fields.

\subsection*{Acknowledgement:}
We register our deep gratitude to Dr. S. H. Ghate for his valuable
guidance. One of the authors (SBS) would like to thank R. Goswami
for many helpful suggestions.

\end{document}